\documentclass[a4paper]{jpconf}
\usepackage{iopams,cite}
\usepackage{graphicx}
\usepackage{amsfonts}
\usepackage{mathrsfs}
\usepackage{epstopdf}
\usepackage{bm}

\newcommand{\be}{\begin{equation}}
\newcommand{\ee}{\end{equation}}
\newcommand{\bea}{\begin{eqnarray}}
\newcommand{\eea}{\end{eqnarray}}

\begin{document}

\title{Resonant driving of a nonlinear Hamiltonian system}

\author{Carlo Palmisano, Gianpiero Gervino}
\address{Dipartimento di Fisica, Universit\`a di Torino, Via P.~Giuria, I--10125 Torino}

\author{Massimo Balma, Dorina Devona}
\address{SELEX Galileo, San Maurizio Canavese (TO), Italy}

\author{Sandro Wimberger}
\address{Institut f\"ur Theoretische Physik and Center for Quantum Dynamics, 
         Universit\"at Heidelberg, D--69120 Heidelberg}
\ead{s.wimberger@thphys.uni-heidelberg.de}

\begin{abstract}
As a proof of principle, we show how a classical nonlinear
Hamiltonian system can be driven resonantly over reasonably long times by appropriately
shaped pulses. To keep the parameter space reasonably small,
we limit ourselves to a driving force which consists of
periodic pulses additionally modulated by a sinusoidal function.
The main observables are the average increase of kinetic
energy and of the action variable (of the non-driven system) with time.
Applications of our scheme aim for driving high frequencies of a nonlinear system
with a fixed modulation signal.
\end{abstract}


\section{Introduction}
\label{intro}\medskip

Driving a nonlinear Hamiltonian system with a fixed form of drive is never effective over long times. The local intrinsic frequency of the system changes with increasing energy, or more generally, with the increasing action of the system. This is the very definition of nonlinearity \cite{LL}. Consequently, to keep the driving in resonance with the system the external frequencies must be gradually adapted. 

In this paper we show that a nonlinear system can nevertheless be pumped with energy from an external drive even if the latter is fixed. The condition for our approach to work is that the driven system allows for global transport in phase space. A very rough, yet effective criterion for the onset of chaos in that sense has been introduced by Chirikov and is by now known as Chirikov (or also resonance overlap) criterion \cite{chirikov}. The overlapping of nonlinear resonances in phase space is the origin of dynamical stochasticity in classical Hamiltonian systems \cite{LL}. In section \ref{sec:2} we introduce a model system which was studied before in reference \cite{levan}, where the effectiveness of the overlap criterion was theoretically predicted. The following section \ref{sec:3} reports major results from ample numerical investigations of the model, which verify the predictions of \cite{levan}. Section \ref{sec:4} concludes the paper.

\begin{figure}[t]
\includegraphics[width=17pc]{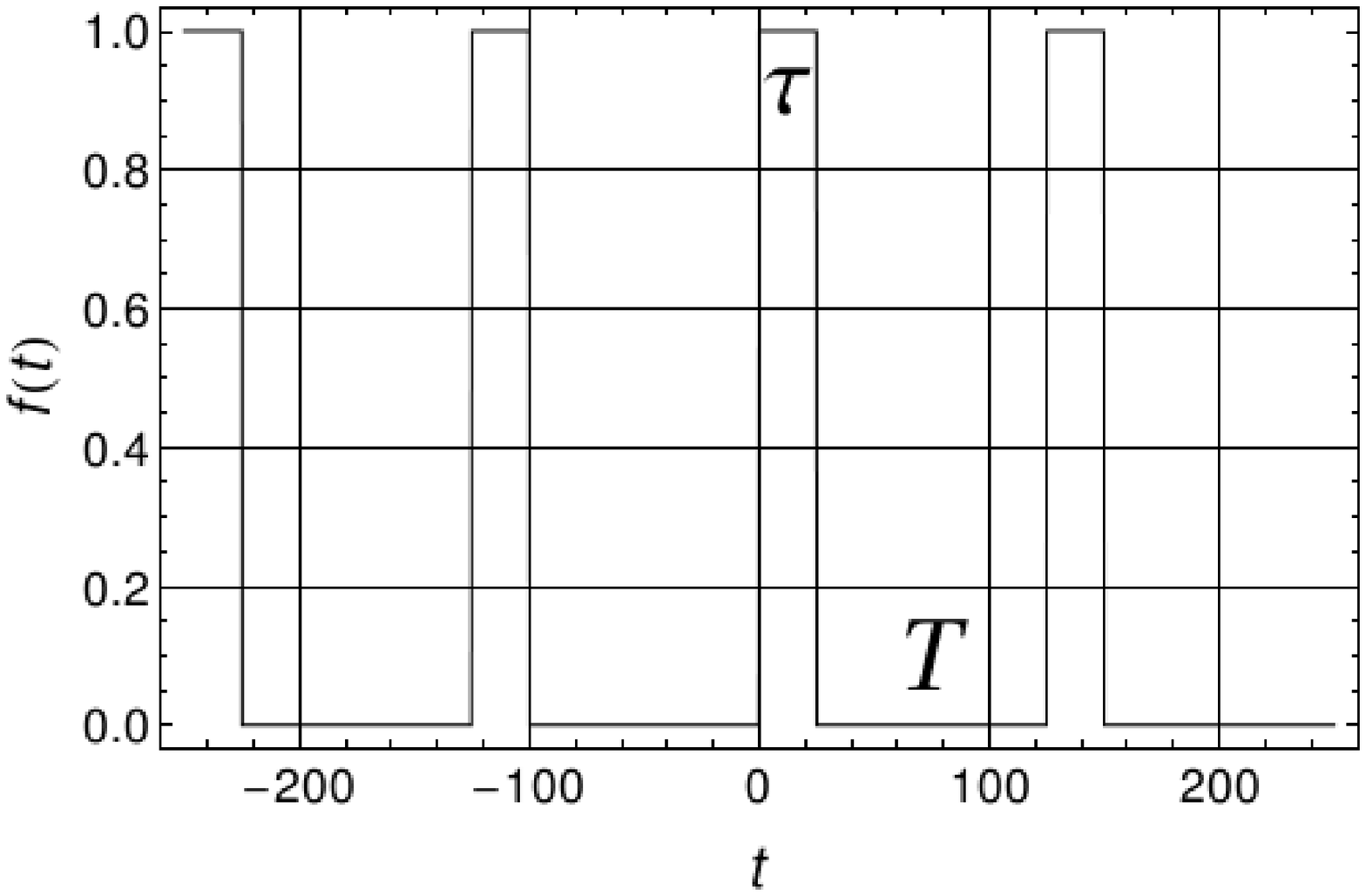}\hspace{2pc}
\begin{minipage}[b]{18pc}
\caption{\label{fig:1} Upper panel:
Form of the pulse train $f(t)$ which is periodic with period $T+\tau$ and the single pulses have a width $\tau$.
Lower panel: Modulation signal $\epsilon V_0 f(t) \sin(\omega t)$. The scales are enhanced for better visibility compared to the parameters chosen latter.
}
\end{minipage}
\includegraphics[width=18pc]{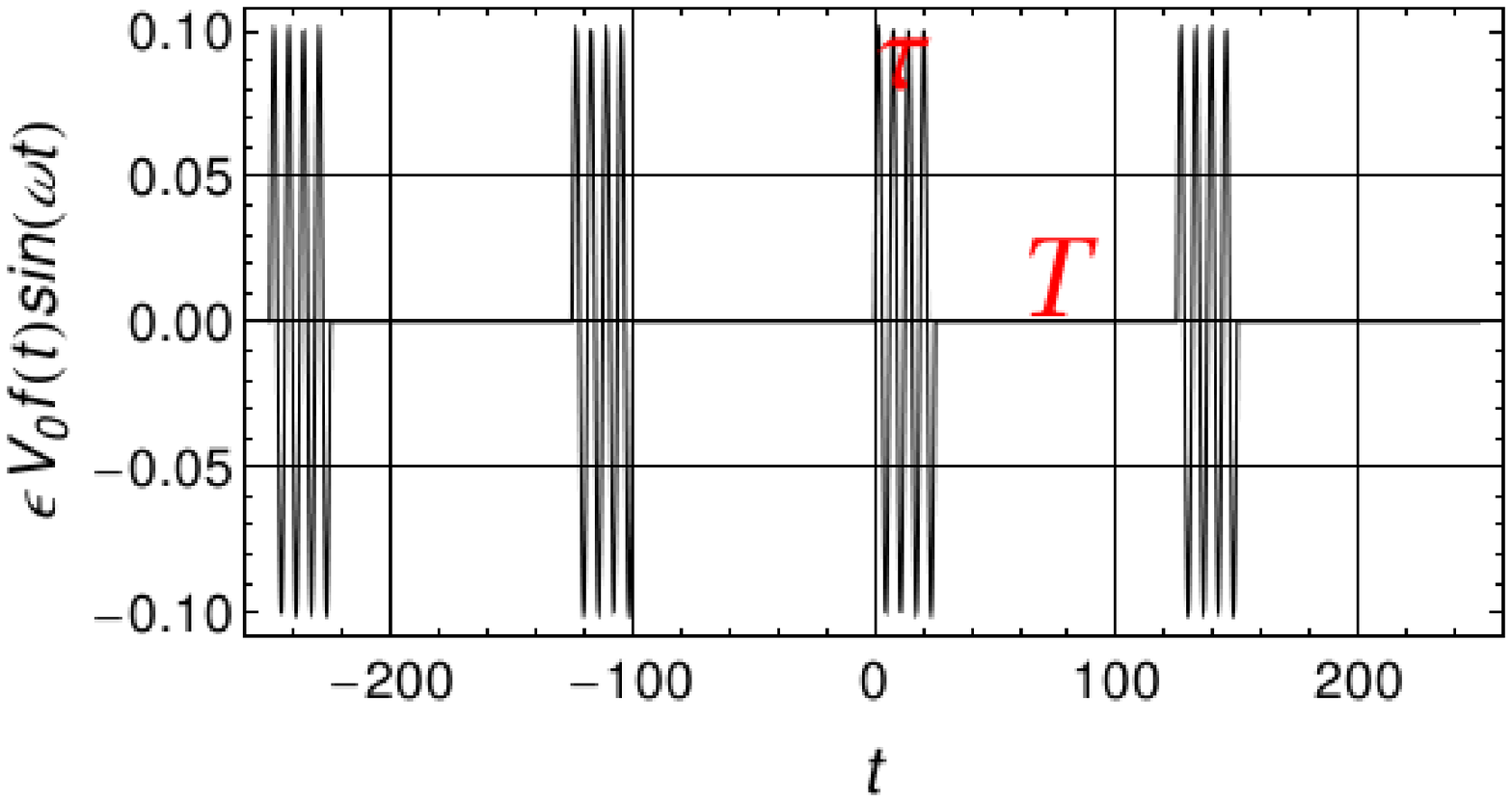}\hspace{2pc}
\end{figure}

\section{Nonlinear model system}
\label{sec:2}\medskip

Throughout we will stick to the following model for a classical Hamiltonian system, given by an harmonic oscillator $H_0$, a quartic nonlinear potential $H_{\mathrm{nl}}$, and a periodic driving $H_{\mathrm{f}}$:
\begin{equation}
H(x,p,t) = H_0(x,p) + H_{\mathrm{nl}}(x) + H_{\mathrm{f}}(x,t) \,,
\label{eq:1}
\end{equation}
with the following parts
\begin{eqnarray}
H_0(x,p) &=& \frac{p^2}{2m}+\frac{1}{2}m\omega_0^2 x^2 \,, \\
H_{\mathrm{nl}}(x) &=& \frac{1}{4}mbx^4 \,, \\
H_{\mathrm{f}}(x,t) &=& -\epsilon V_0 x f(t)\sin(\omega t)\,.
\label{eq:2}
\end{eqnarray}
Similar models (also known as Lorentz models) have been studied over many decades, in particular in the context of nonlinear optics \cite{boyd}. Instead of integrating directly Newton's equation corresponding to the Hamiltonian (\ref{eq:1}) in the variables $(x,p)$, we follow reference \cite{levan} and rewrite the problem in action-angle variables of the harmonic oscillator $(\theta,J)$ which are connected to the previous ones by the canonical transformation
\begin{eqnarray}
x(\theta, J)&=&\sqrt{\frac{2J}{m\omega_0}}\sin\theta \label{eqs:xJtheta} \,, \\
p(\theta, J)&=&\sqrt{2Jm\omega_0}\cos\theta \,.
\label{eqs:3}
\end{eqnarray}
The advantage of using the action-angle variables lies not only in their use in perturbation theory as exercised in \cite{levan}, but for us, first and foremost, in the fact that we are interested in transport in phase space. The latter is characterized best by the temporal evolution of the action $J$ of the ``unperturbed system''. In the new variables we obtain the new Hamiltonian
\begin{equation}
H(x,p,t) \longrightarrow H'(\theta, J, t) = J \omega_0+
\frac{J^2 b \sin^4\theta}{m\omega_0^2} - \epsilon V_0 \sqrt{\frac{2J}{m\omega_0}}\sin\theta f(t)\sin(\omega t) \,.
\label{eq:4}
\end{equation}
The corresponding equations of motion for $(\theta,J)$ are then simply given by Hamilton's equations:
\begin{eqnarray}
\frac{\mathrm{d}\theta}{\mathrm{d}t}(t)&=&\frac{\partial H'}{\partial J}(\theta,J,t) \nonumber \\
                                       &=&\omega_0
                                        +2\frac{J b \sin^4\theta}{m\omega_0^2}
                                        -\epsilon V_0\frac{1}{2\sqrt{J}}
                                        \sqrt{\frac{2}{m\omega_0}}\sin\theta f(t)\sin(\omega t) \\
\frac{\mathrm{d}J}{\mathrm{d}t}(t)&=&-\frac{\partial H'}{\partial \theta}(\theta,J,t) \nonumber \\
                                  &=&-4\frac{J^2 b \sin^3\theta\cos\theta}{m\omega_0^2}
                                   +\epsilon V_0 \sqrt{\frac{2J}{m\omega_0}}\cos\theta f(t)\sin(\omega t)\,.
\label{eqs:5}
\end{eqnarray}
In the next section we present results obtained by integrating these equations of motion over time. Before we motivate our choice of the driving signal given by $H_{\mathrm{f}}$. It essentially contains two frequency scales, one given by the sinusoidal part oscillating with frequency $\omega$ and a second one determined by a period $T+\tau$. The detailed form
of the drive is best seen pictorially in Fig.~\ref{fig:1}. Now we fix this form of the external driving and try to investigate the behavior of the time-averaged action $\langle J(t)\rangle_t=1/t \int_0^t J(t') dt'$ for different parameters of the model.

\section{Numerical results}
\label{sec:3}\medskip

For the following computations, we fix the initial conditions $x(0)=0, mx'(0)=p(0)=-2$ and the parameters
\begin{equation}
m=1,  \omega_0 = 20, \tau=1, \epsilon  = 0.5, V_0 =100, T=10\,\tau,  \omega=\frac{16\pi}{\tau}  \,.
\label{eq:6}
\end{equation}
Units have been suppressed throughout (and maybe easily reintroduced in case having in mind the meaning of all parameters). This leaves us only one control parameter, the effective nonlinearity $b$, which characterizes the nonlinear response of the system with respect to the external drive. 

We do not repeat the analytical estimates for the widths of corresponding nonlinear resonances in the phase space spanned by the variables $(\theta, J)$. Reference \cite{levan} derives from them a critical parameter $K$ which separates motion without overlapping resonances, for $K<1$, from the one with overlapping resonances, implying chaotic transport for $K>1$. Written in our parameters, we obtain from \cite{levan}
\begin{equation}
K (J)= \frac{\tau+T}{2\pi}\sqrt{\frac{\partial H'_{\mathrm{nl}}(J)}{\partial J}
\,\frac{\epsilon}{4}\,\frac{2\pi}{\tau+T}V_0\sqrt{\frac{2J}{m\omega_0}}}>1\,.
\label{eq:7}
\end{equation}
$K$ is time-dependent through the dependence on the action variable $J$. Its initial value at time $t=0$ indeed determines the increase of the mean action as the following figures show.

Figs. \ref{fig:2} and \ref{fig:3} compare the evolution of the actions for various nonlinearities $b=b_1=3200$ (left panel in Fig. \ref{fig:2}), $b=b_1/2$ (left panel of Fig. \ref{fig:3}), and $b=b_1/4$ (right panel of Fig. \ref{fig:3}). These parameters imply for Eq.~(\ref{eq:7}) $K(J(t=0))=1.44$, $1.02$, and $0.72$, respectively. The shown results confirm in practice that the initial value of the parameter $K$ indeed determines whether transport in action space occurs (see the left panels of Figs. \ref{fig:2} and \ref{fig:3}) or is suppressed by transport barriers in phase space (see the right panel of Fig. \ref{fig:3}).

\begin{figure}[ht]
\includegraphics[width=18pc]{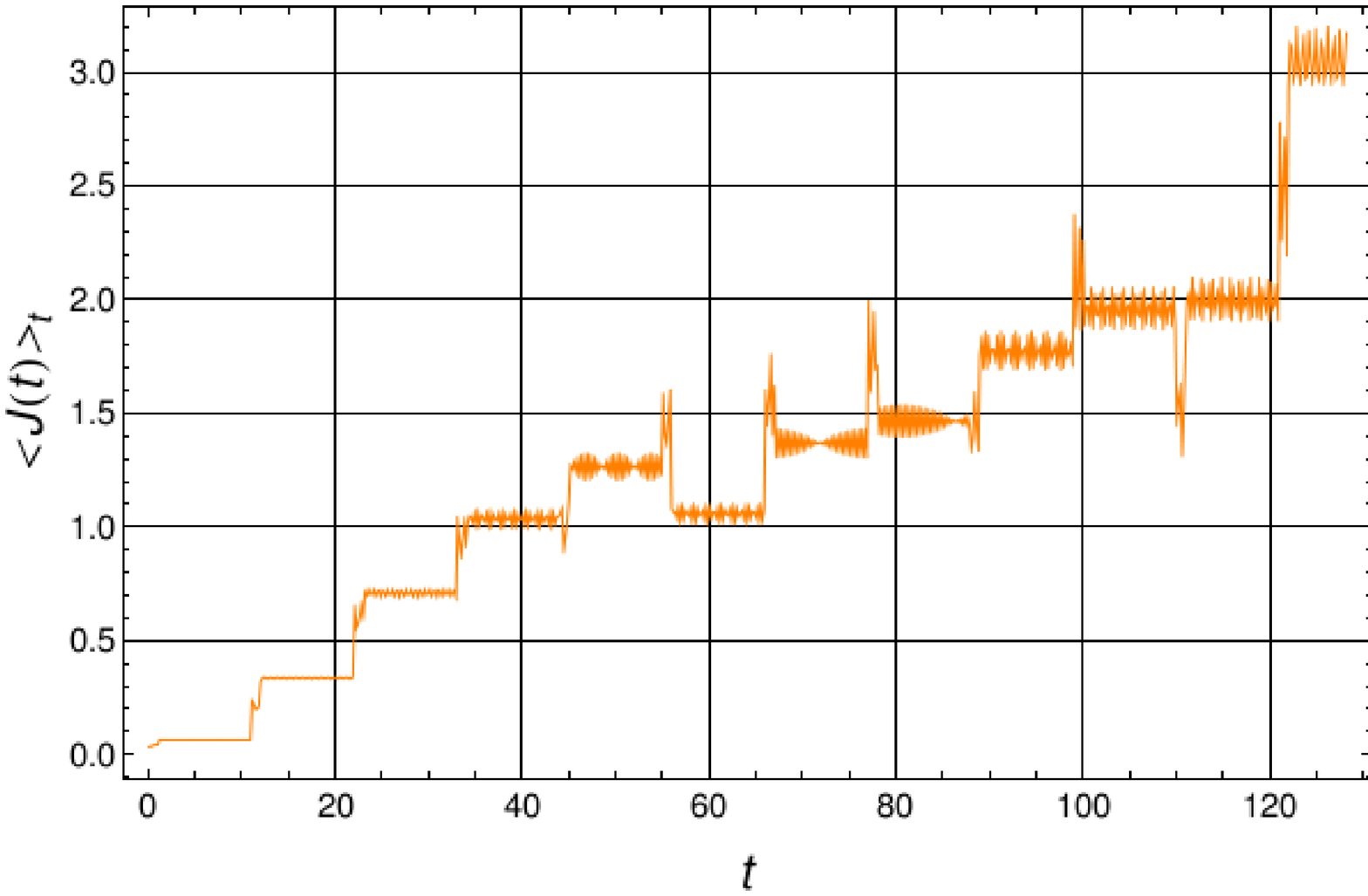}\hspace{1pc}
\includegraphics[width=18pc]{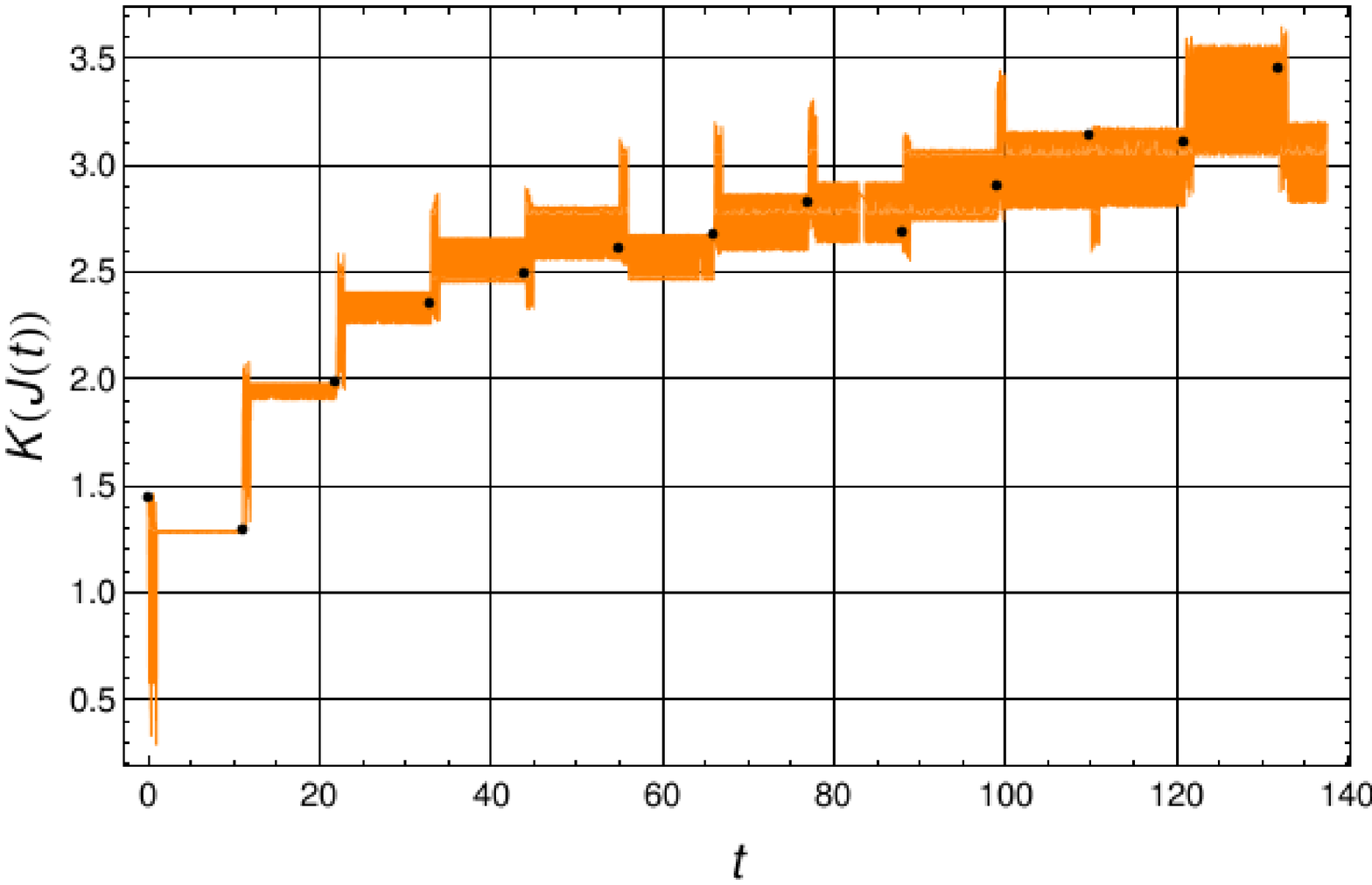}
\caption{\label{fig:2} 
Time averaged action $\langle J(t)\rangle_t$ (left panel) and critical parameter $K$ (right panel) vs. time $t$ for $b=3200$, implying $K(J(t=0))=1.44$. The overall increase of both quantities with time is clearly visible confirming the theoretical predictions.
}
\end{figure}

\begin{figure}[ht]
\includegraphics[width=18pc]{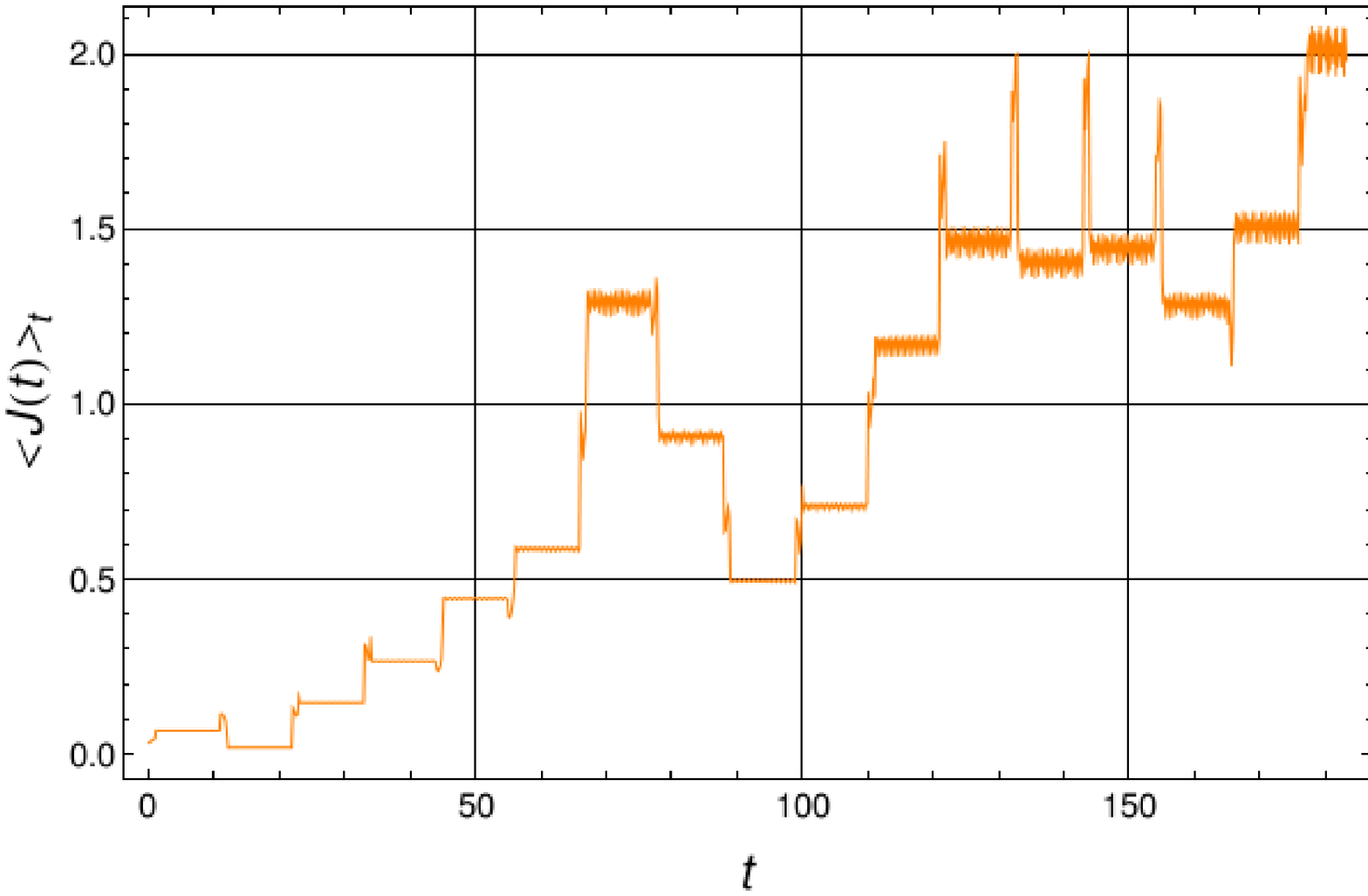}\hspace{1pc}
\includegraphics[width=18pc]{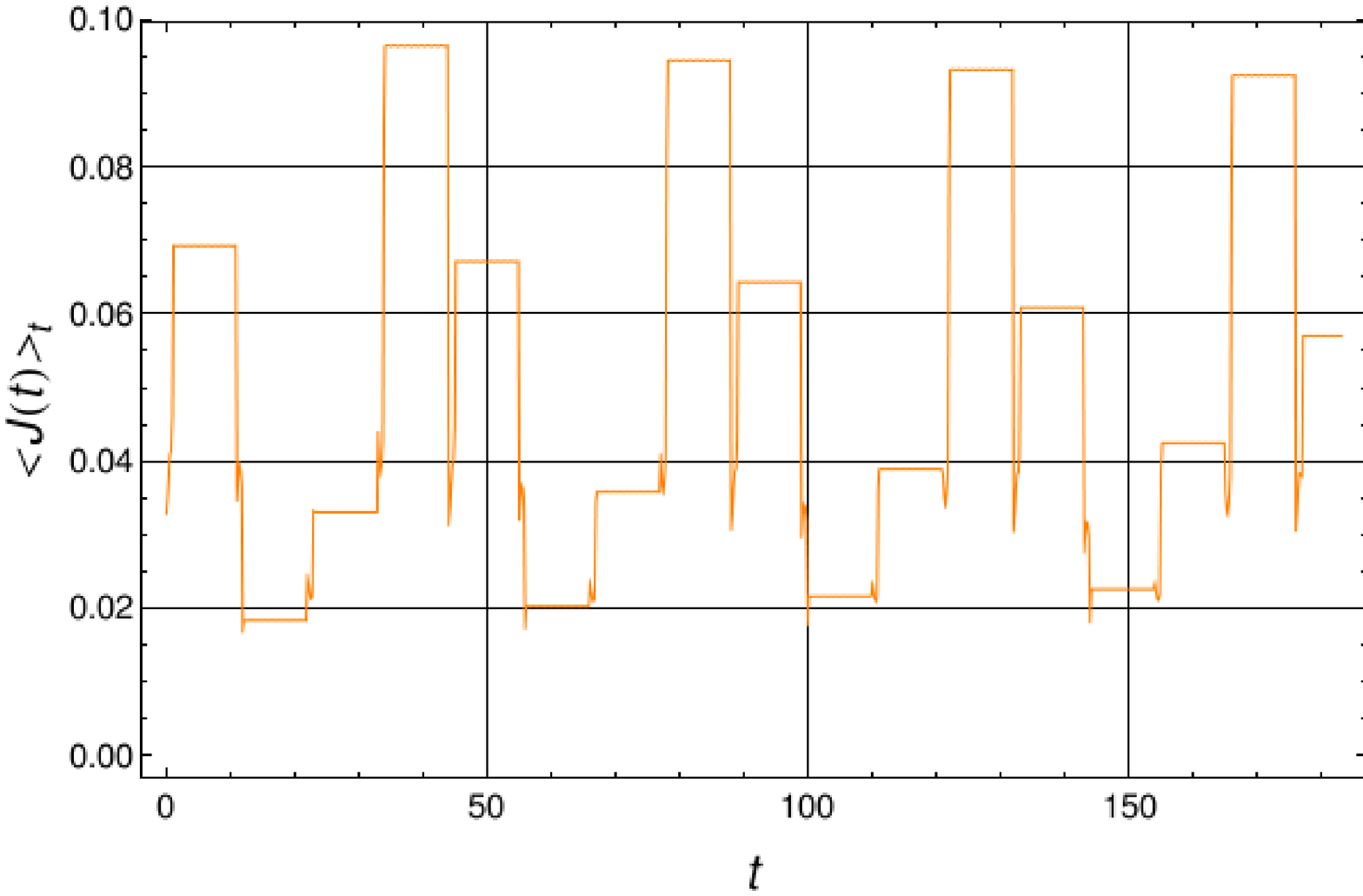}
\caption{\label{fig:3} 
Time averaged actions $\langle J(t)\rangle_t$ vs. time $t$ for two sets of parameters: $b=1600$ and $K(J(t=0))=1.02$ (left panel) and $b=800$ and $K(J(t=0))=0.72$ (right panel). While the left plot is similar to the previous Fig. \ref{fig:2}, the right panel shows a case below the criterion for transport, and hence the action just oscillates around small values.}
\end{figure}

\section{Conclusions} 
\label{sec:4}\medskip

We studied a nonlinear model system for controlling the temporal evolution by an appropriate external driving. 
Exploiting the nonlinear response of the system to the drive, we confirmed that resonant driving is possible as long
as there are no transport barriers hindering diffusion in phase space. The transporting and the non-transporting regimes are separated by a critical effective nonlinearity which determine the value of a control parameter introduced in \cite{levan}.

Our results give a proof of principle of the applied method. In the future we want to extend the idea to nonlinear systems in the presence of additional noise, as suggested in the recent reference \cite{levan-2}.

\ack
S W and G G thank with pleasure the organiser Thomas Elze for his kind invitation to DICE2012.
We acknowledge also the help of the Heidelberg Graduate School of Fundamental Physics (DFG grant: GSC 129/1).

\end{document}